\documentclass[superscriptaddress,prd,reprint,showpacs]{revtex4-1}

\usepackage{amsmath,empheq}
\usepackage{amsfonts}
\usepackage{amsthm}
\usepackage{amssymb}
\usepackage{graphicx}
\usepackage{multirow}
\usepackage{hyperref}
\usepackage{soul}
\usepackage{url}
\usepackage{color}
\usepackage{bm}
\usepackage{cleveref}
\usepackage[export]{adjustbox}

\usepackage{chngcntr}
\counterwithout{table}{section}

\graphicspath{ {./images/} }

\newcommand{\ncd}{\newcommand}
\ncd{\mrm}    {\mathrm}
\ncd{\beq} {\begin{equation}}
  \ncd{\eeq} {\end{equation}}

\begin{document}
\title{Ergodicity of one-dimensional systems coupled to the logistic thermostat}
\date{\today}

\author{Diego Tapias}
\email{diego.tapias@nucleares.unam.mx}
\affiliation{Departamento de F\'isica, Facultad de Ciencias, Universidad Nacional Aut\'onoma de M\'exico, Ciudad Universitaria, Ciudad de M\'exico 04510, Mexico}

\author{Alessandro Bravetti}
\email{alessandro.bravetti@iimas.unam.mx}
\affiliation{Instituto de Investigaciones en Matem\'aticas Aplicadas y en Sistemas, Universidad Nacional Aut\'onoma de M\'exico,
  Ciudad Universitaria, Ciudad de M\'exico 04510, Mexico}

\author{David P. Sanders}
\email{dpsanders@ciencias.unam.mx}
\affiliation{Departamento de F\'isica, Facultad de Ciencias, Universidad Nacional Aut\'onoma de M\'exico, Ciudad Universitaria, Ciudad de M\'exico 04510, Mexico}
\affiliation{Computer Science and Artificial Intelligence Laboratory,  Massachusetts Institute of Technology, 77 Massachusetts Avenue, Cambridge, MA 02139, USA}

\begin{abstract}
We analyze the ergodicity of three one-dimensional Hamiltonian systems, with  harmonic, quartic and Mexican-hat potentials, 
coupled to the logistic thermostat.
As criteria for ergodicity we employ: 
the independence of the Lyapunov spectrum with respect to initial conditions;
the absence of visual ``holes'' in two-dimensional Poincar\'e sections;
 the agreement between the histograms in each variable and the theoretical marginal distributions; and the convergence of the global joint distribution to the theoretical one, as measured by the Hellinger distance. 
Taking a large number of random initial conditions, for certain parameter values of the thermostat we find no indication of regular trajectories and 
show that the time distribution converges to the ensemble one for an arbitrarily long trajectory for all the systems considered.
Our results thus provide a robust numerical indication that the logistic thermostat can serve  
 as a single one-parameter thermostat for stiff one-dimensional systems.
\end{abstract}

\maketitle

\section{Introduction}

The introduction by Nos\'e and Hoover of deterministic equations of motion consistent with the canonical ensemble allowed to make
a connection between microscopic and macroscopic descriptions for ensembles different from the microcanonical~\cite{nose1984canonical, hoover1985canonical}. 
However, there is a practical limitation that impedes the use of the Nos\'e--Hoover equations for a given system, namely ergodicity. 
Roughly speaking, 
a system is \emph{ergodic} if for almost any trajectory, taking long-time averages is equivalent to taking
 ensemble averages~\cite{oliveira2007ergodic, aleksandr1949mathematical}. 
For the majority of physical systems, ergodicity can be tested only through numerical experiments. 

The Nos\'e--Hoover thermostat fails to be ergodic for a one--dimensional harmonic oscillator \cite{hoover1985canonical}. 
Therefore, various alternative schemes have been proposed to simulate
a harmonic oscillator in the canonical ensemble \cite{kusnezov1990canonical, tuckerman1992chains, hoover1996kinetic,branka2003generalization, sergi2010bulgac, hoover2016nonequilibrium}, 
some of which seem to be ergodic, {in the sense that they pass a series of different numerical tests designed to detect this property}. 
Among the ergodic schemes, the ``0532'' thermostat is the only one that requires the addition of a single thermostatting force~\cite{hoover2016nonequilibrium}
(see also the discussion in~\cite{ramshaw2015general}).

The ``0532'' model was inspired by 
the observation that a cubic thermostat force enhances ergodicity with respect to the linear (Nos\'e--Hoover)
one~\cite{kusnezov1990canonical, hoover2016nonequilibrium}. 
Thus the authors in~\cite{hoover2016nonequilibrium} started with a general parametric 
three-dimensional 
dynamical system with a cubic friction force, designed to control directly the first three even moments of {the momentum} $p$.
They then adjusted the parameters for the case of a harmonic potential, using a $\chi^2$ test, by imposing that the joint probability distribution be Gaussian in the three variables.

The method described in the last paragraph can be extended in principle to more general one-dimensional potentials.
However, there are two major drawbacks. First, one has to repeat the $\chi^{2}$ test for each potential, which is a computationally 
demanding task. Second, the form of the parametric equations to be tested may depend on the potential of the system to be 
thermostatted and thus the idea of generality behind the Nos\'e-Hoover equations is lost.
Furthermore, the analysis in~\cite{hoover2016singly} has shown that this thermostat works well for the one-dimensional harmonic oscillator, but not for the quartic potential.

For these reasons it is relevant to ask if there is a general scheme depending just on the addition of a single thermostatting force that allows the generation of 
a large family of
ergodic singly-thermostatted one-dimensional systems (ST1DS). 
This is the challenge of the 2016 Ian Snook prize~\cite{hoover2016singly} and the subject of this work.

We start from an algorithm to generate the equations of motion known as Density Dynamics~\cite{fukuda2002tsallis}. 
Combining this scheme with the \emph{logistic thermostat} introduced previously \cite{bravtap2016, tapias2016geometric} by two of the present authors, we generate a set of ST1DS  
{for different potentials} and we show  {that such systems pass all the numerical tests for ergodicity}.
The advantage of the Density Dynamics formalism is that the equations of motion are the same in form for any Hamiltonian system, 
thus retaining the spirit of generality of Nos\'e and Hoover \cite{fukuda2002tsallis}.
The superiority of the logistic thermostat comes from the fact that the thermostatting force is highly nonlinear, thus enhancing the ergodicity of the dynamics.
Additionally, we show that the equations of motion that we obtain are time-reversible. All these aspects make the logistic thermostat appealing from both a practical and a theoretical perspective.

The structure of the paper is as follows. In section~\ref{dd}, we give an introduction to the Density Dynamics formalism and present the logistic thermostat. 
In section~\ref{results}, we present the numerical methods used to study ergodicity, together with the results obtained. 
Finally, in section~\ref{conclusions} we summarize our results and present the conclusions.

\section{Density Dynamics}
\label{dd}

The Density Dynamics (DD) method was introduced by Fukuda and Nakamura, inspired by the Nos\'e--Hoover equations of motion \cite{fukuda2002tsallis}. 
Afterwards, the same method was re-derived by Bravetti and Tapias, starting from a dynamics based on a generalization of 
Hamilton's equations~\cite{bravetti2015liouville,tapias2016geometric, bravtap2016}. 

The DD method provides an algorithm for the generation of a set of equations in a 
$(2n+1)$-dimensional space consistent with 
a prescribed probability distribution ($n$ being the degrees of freedom of the physical system). 
For a general description of the method we refer to~\cite{fukuda2002tsallis, tapias2016geometric, bravtap2016}. In this section we present its application to ST1DS.

Let $n=1$ and consider the $3$-dimensional extended phase space with coordinates $(q, p, \zeta)$. 
A  one-dimensional Hamiltonian system coupled to a thermostat is expected to present a canonical probability distribution in $(q,p)$. 
So, the invariant distribution to be generated in $(q,p,\zeta)$ is of the form 
	\beq
	\rho(q,p,\zeta) = \frac{{\rm e}^{-\beta H(q,p)}}{\cal Z} f(\zeta) \, ,
	\label{probdensity}
	\eeq
where $\mathcal{Z}$ is a normalization constant and $f(\zeta)$ is a 1-dimensional probability distribution in $\zeta$, i.e.~$f(\zeta)$ is a 
strictly positive, smooth, integrable function with support in $\mathbb{R}$. 
According to the DD prescription, the equations of motion consistent with the probability density \eqref{probdensity} are
	\begin{empheq}{align}
	  \label{canon1}
	  & \dot{q} =  \frac{\partial H(q,p)}{\partial p}  \, , \\
	  \label{canon2}
	  & \dot{p} =  - \frac{\partial H(q,p)}{\partial q}  + \frac{f'(\zeta)}{\beta f(\zeta)}p 	\,, \\
	  \label{canon3}		
	  & \dot{\zeta} =  \frac{\partial H(q,p)}{\partial p} p - \frac{1}{\beta} \, .
	\end{empheq}

Consistency between the field  $v = (\dot{q}, \dot{p}, \dot{\zeta})$ and the distribution \eqref{probdensity}  means that the Liouville equation is satisfied for this pair, i.e.

\begin{widetext}
\begin{eqnarray}
div(\rho v) &=& \nabla\cdot(\rho v) = (\nabla \rho) \cdot v + \rho (\nabla \cdot v)   =  \frac{\partial \rho}{\partial q}\dot{q} + \frac{\partial \rho}{\partial p}\dot{p} + \frac{\partial \rho}{\partial \zeta}\dot{\zeta} + \rho \left( \frac{\partial \dot{q}}{\partial q} + \frac{\partial \dot{p}}{\partial p} + \frac{\partial \dot{\zeta}}{\partial \zeta} \right) \notag \\
&=& \rho\left( -\beta \frac{\partial H}{\partial  q}\frac{\partial H}{\partial  p} 
- \beta\frac{\partial H}{\partial  p} \left(-\frac{\partial H}{\partial  q} + \frac{f'}{\beta f}\,p \right) 
+ \frac{f'}{f} \left(\frac{\partial H}{\partial  p}p - \frac{1}{\beta} \right)\right) + \rho \left(\frac{\partial^2 H}{\partial  p \partial q} - \frac{\partial^2 H}{\partial  q \partial p} + \frac{f'}{\beta f} \right) = 0\,.
\end{eqnarray}
\end{widetext}
Naturally, this proof extends directly to systems with more degrees of freedom. 

\subsection{The logistic thermostat}

The set of equations \eqref{canon1}--\eqref{canon3} depends on the probability distribution chosen for the extended variable $f(\zeta)$, associated with the effect of the thermal reservoir. 
By choosing a Gaussian distribution with variance $Q$ and mean $0$, we recover the 
time-reversible Nos\'e--Hoover equations of motion. 
These dynamical equations modify the structure of Hamilton's equations by adding a linear friction term that obeys a feedback equation 
that controls the kinetic energy~\cite{hoover1985canonical}. 
For the same system, one can consider different distributions $f(\zeta)$. For instance, a Gaussian distribution for $\zeta^2$ introduces a cubic friction term, which considerably improves ergodicity~\cite{kusnezov1990canonical, fukuda2002tsallis, hoover2016ergodicity}. 

Following the observation that nonlinearity enhances ergodicity, we choose $f(\zeta)$ to be a logistic distribution:
	\begin{equation}
	f_{\rm logistic}(\zeta) = \frac{ {\rm e}^ {\frac{\zeta - \mu}{Q}} }{ Q(1 + {\rm e}^{\frac{\zeta - \mu}{Q}})^2 } \, ,
	\label{logisticdistribution}
	\end{equation}
where $\mu$ is the mean of the distribution and the variance is $Q^2\pi^2/3$. 
We call this choice the \emph{logistic thermostat} and refer to $Q$ as the ``mass'' associated with the thermostat, 
using the same terminology as for the Nos\'e--Hoover case \cite{hunenberger2005thermostat}. 

In our previous works we used the logistic thermostat with the choice of the parameters $Q = 1$ and $\mu = 2$ and we showed that this
 is a suitable choice to perform molecular dynamics simulations~\cite{bravtap2016, tapias2016geometric}. 
 However, these particular values make the resulting dynamical system not time--reversible, 
 which is an important property for a dynamical model that aims to simulate equilibrium. 
 Here we fix this issue by suggesting a different parameter choice. Choosing  
 $\mu = 0$, we see that $f(\zeta)$ becomes an even function and it follows that the corresponding equations of motion 
\begin{empheq}{align}
  \label{logistic1}
  & \dot{q} =  \frac{\partial H(q,p)}{\partial p}  \, , \\
  \label{logistic2}
  & \dot{p} =  - \frac{\partial H(q,p)}{\partial q}  + \frac{1-{\rm e}^{\frac{\zeta}{Q}}}{\beta Q(1+{\rm e}^{\frac{\zeta}{Q}}) }\, p 	\,, \\
  \label{logistic3}		
  & \dot{\zeta} =  \frac{\partial H(q,p)}{\partial p}\, p - \frac{1}{\beta} \, .
\end{empheq}
are time-reversible, i.e.~invariant under the transformation $(q,p,\zeta,t)\rightarrow(q,-p,-\zeta,-t)$.
Equations \eqref{logistic1}--\eqref{logistic3} constitute our system that provides thermostatted dynamics for any one-dimensional Hamiltonian system
encoded in $H(q,p)$.

\section{Numerical {tests} and results}
\label{results}

In this section we numerically test the ergodicity of the system \eqref{logistic1}--\eqref{logistic3} for three Hamiltonian systems with Hamiltonians given by
\beq
H(p,q) = \frac{p^2}{2} + V(q) \, ,
\eeq
with potentials 
\begin{itemize}
\item $V(q) = q^2/2$ (harmonic);
\item $V(q) = q^4/4$ (quartic);
\item $V(q) = -q^2/2 + q^4/4$ (Mexican hat).
\end{itemize}
Throughout {this} section the (inverse) temperature is taken as $\beta = 1.0$. 
The  ``mass'' of the thermostat for the harmonic and quartic systems is $Q = 0.1$, 
whereas for the Mexican hat potential it is $Q = 0.02$. 
These values were chosen on the base of preliminary tests designed to detect violations of ergodicity.
For instance, for $Q=0.1$ in the case of the Mexican hat potential we found $5$ regular trajectories out of 1 million initial conditions, thus indicating 
a possible violation of ergodicity. For such reason the value of $Q$ considered for such potential is different from the one used for the other systems.

{Before proceeding with the numerical analysis, we summarize the relationship between such tests and ergodicity.
In essence, an ergodic {thermostatted} system is expected to present a single chaotic sea of full measure in its extended phase space, 
so that for almost any initial condition in this set, the numerical distribution in time 
converges to the theoretical distribution in the ensemble~\cite{branka2003generalization, patra2015ergodic, patra2015deterministic}. 
The study of the chaotic sea relies on  both the analysis of the Lyapunov spectrum for a large number of initial conditions and on the observation of Poincar{\'e} sections. 
With these tests one checks the independence of the spectrum with respect to the initial condition and discards the presence of islands that would violate the assumption that 
the chaotic sea has full measure.
Then one proceeds to analyse the equivalence between the numerical distribution and the theoretical one. 
For this, one observes the visual agreement between the numerical histograms and the marginal theoretical distributions and checks the mean values of certain 
observables~\cite{kusnezov1990canonical,tuckerman1992chains, fukuda2002tsallis, leimkuhler2010generalized}. 
Recently, stronger tests have been used to analyze the convergence between distributions, based on distances in the distributions space~\cite{leimkuhler2009gentle, patra2014nonergodicity}. 
Here we consider the Hellinger distance~\cite{basu2011statistical, patra2014nonergodicity}.}

\subsection{Lyapunov characteristic exponents}

For a dynamical system, the Lyapunov characterisic exponents (LCEs) are asymptotic measures characterizing the average rate of growth (or shrinking) 
of small perturbations of the solutions \cite{skokos2010lyapunov}. 
The set of LCEs is grouped in the Lyapunov spectrum. 

There are three facts about the Lyapunov spectrum that are relevant for our numerical study: 
if the largest exponent in the spectrum for a given trajectory is greater than zero, then the trajectory is chaotic; 
if the sum of exponents in the spectrum for a given trajectory is equal to zero, then its nearby volume is maintained on average; finally, 
if the spectrum is independent of the initial condition, then the system is ergodic. 

In the following, we report the numerical conditions used and discuss our results;
for a similar study for different thermostat models, see Ref.~\cite{patra2015deterministic}.
We take ten thousand random initial conditions for each system, with a weight given by the logistic distribution in $\zeta$ with mean $\mu=0$ {and $Q$ chosen according to the potential}, as specified above,
and by the normal distribution in $p$ and $q$, with mean $0$ and variance $1$ for each variable. 
We follow the procedure of Bennetin et al. \cite{bennetin1980lyapunov, skokos2010lyapunov}
to calculate the Lyapunov spectrum by setting up the variational equations associated with the system  \eqref{logistic1}--\eqref{logistic3} 
and solving them together with the original system
for each initial condition, using a fourth-order Runge--Kutta integrator with a step size of $0.005$ and $10^7$ time steps. 

The relevant results regarding the Lyapunov spectra for each case are reported in table \ref{lyapexpstable}. 
	 \begin{table}[h!]
	\resizebox{\columnwidth}{!}{
	\begin{tabular}{|c|c|c|c|}
	\hline 
	 & $ \lambda_1 $ & $ \lambda_2 $ & $\lambda_3 $ \\
	\hline 
	  Harmonic & $0.281 \pm 2\times10^{-3}$ & $0.000 \pm 3\times10^{-5}$ &
	  $-0.281 \pm 2\times10^{-3}$ \\ 
	\hline 
	Quartic  & $0.243  \pm 2\times10^{-3}$ & $0.000 \pm 4\times10^{-5}$ &
	$-0.243 \pm 2\times10^{-3}$ \\  
	\hline 
	Mexican hat & $0.385 \pm 7\times10^{-3}$ & $0.000 \pm 7 \times 10^{-3}$ &\
	$-0.386 \pm 7\times10^{-3}$ \\ 
	\hline 	
	\end{tabular} 
	}
	\caption{Mean Lyapunov characteristic exponents, {estimated with}
	$10000$ different random initial conditions. The errors are  standard deviations.}\label{lyapexpstable}
	\end{table}
With this test we deduce that the systems are chaotic and that the exponents within each spectrum add to zero, 
thus characterizing an equilibrium system (zero average contraction of volume in the extended phase space). 
{Furthermore, the small relative value of the standard deviation {suggests} the independence of the spectra with respect to the initial condition. 

{We now proceed to analyze in depth this property. For that, we consider one million initial conditions and integrate 
the equations of motion needed to obtain the largest LCE for a short time, but sufficiently long to discriminate between a regular and a chaotic trajectory, which we estimate as $500$ times the Lyapunov time (inverse of the largest Lyapunov exponent \cite{skokos2010lyapunov}). 
Then we check the consistency between the exponent obtained and the expected one as given in table~\ref{lyapexpstable}. 
When a possible regular trajectory is detected via an anomalously low value of the largest Lyapunov exponent, the equations are integrated for a longer time. 
We find that for the three systems considered the spectrum 
is independent of the initial condition.}}

\subsection{Poincar\'e sections}

The second test of ergodicity is based on Poincar\'e sections for a very long trajectory. The visual observation of ``holes'' in these sections is an indication
of the lack of ergodicity \cite{hoover2016ergodicity}. 

\begin{figure}[h!]
  \centering
  \includegraphics[width=\columnwidth]{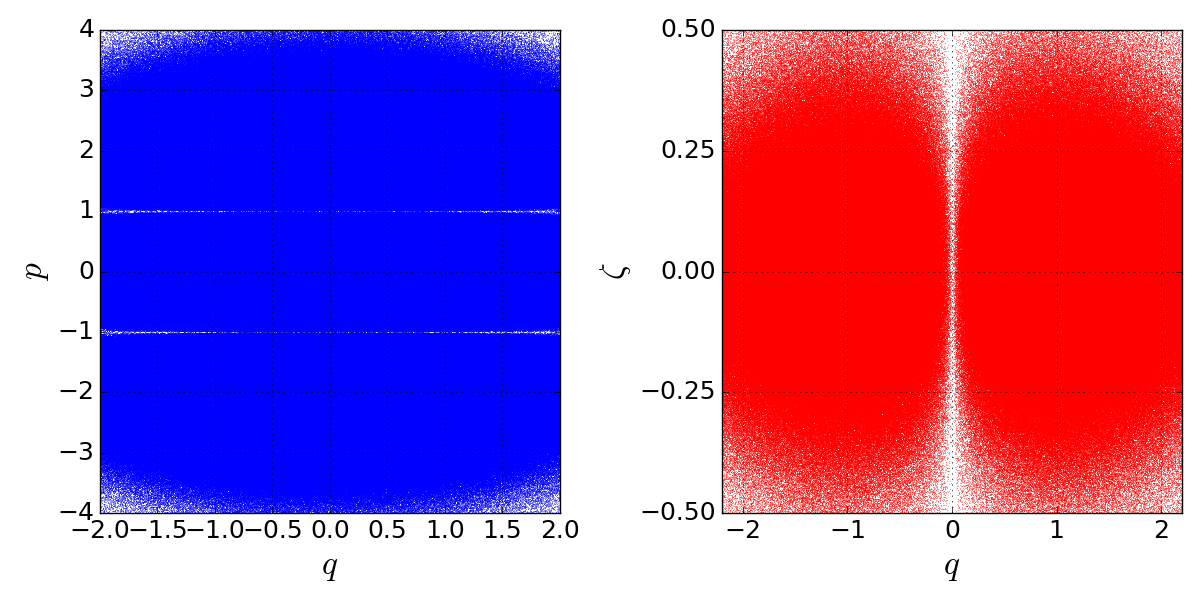}
  \caption{Poincar\'e sections for the harmonic potential. Around $3 \times 10^7$ crossings are shown for the section $\zeta = 0$ (left)
  {and} $3 \times 10^6$ crossings for the section $p = 0$ (right). Additionally, the nullcline lines $p = \pm 1$ (left) and $q = 0$ (right) are observed. }
  \label{poincareharmonic}
\end{figure}

\begin{figure}[h!]
  \centering
  \includegraphics[width=\columnwidth]{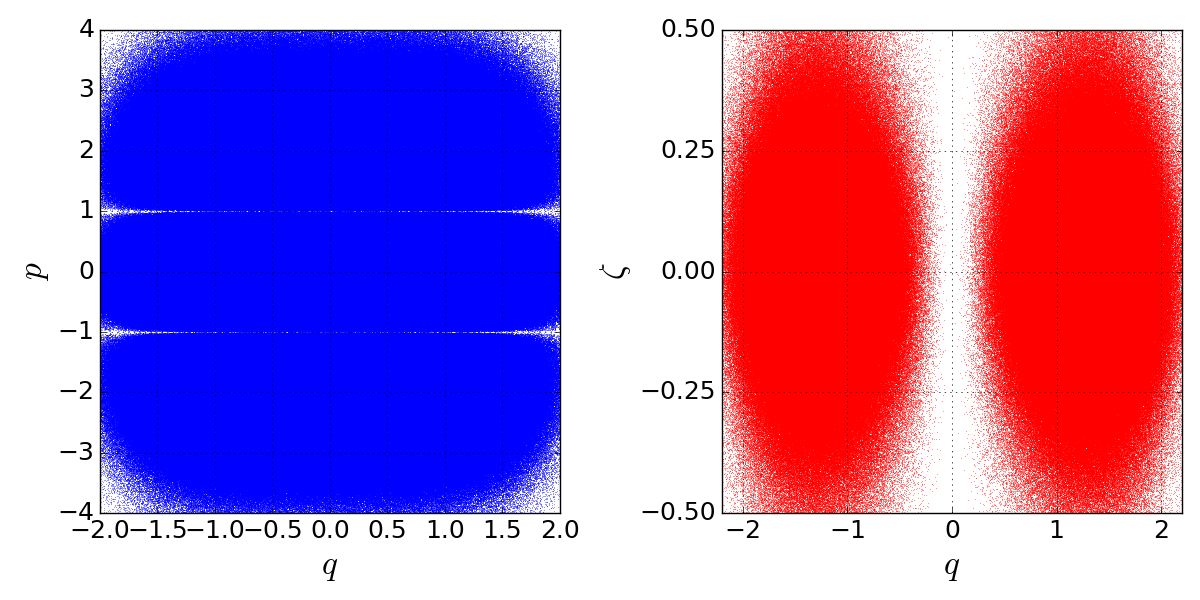}
  \caption{Poincar\'e sections for the quartic potential. $3 \times 10^7$ crossings are shown for the section $\zeta = 0$ (left)
  {and} $3 \times 10^6$ crossings for the section $p = 0$ (right). Additionally, the nullcline lines $p = \pm 1$ (left) and $q = 0$ (right) are observed.}
  \label{poincarequartic}
\end{figure}

\begin{figure}[h!]
  \centering
  \includegraphics[width=\columnwidth]{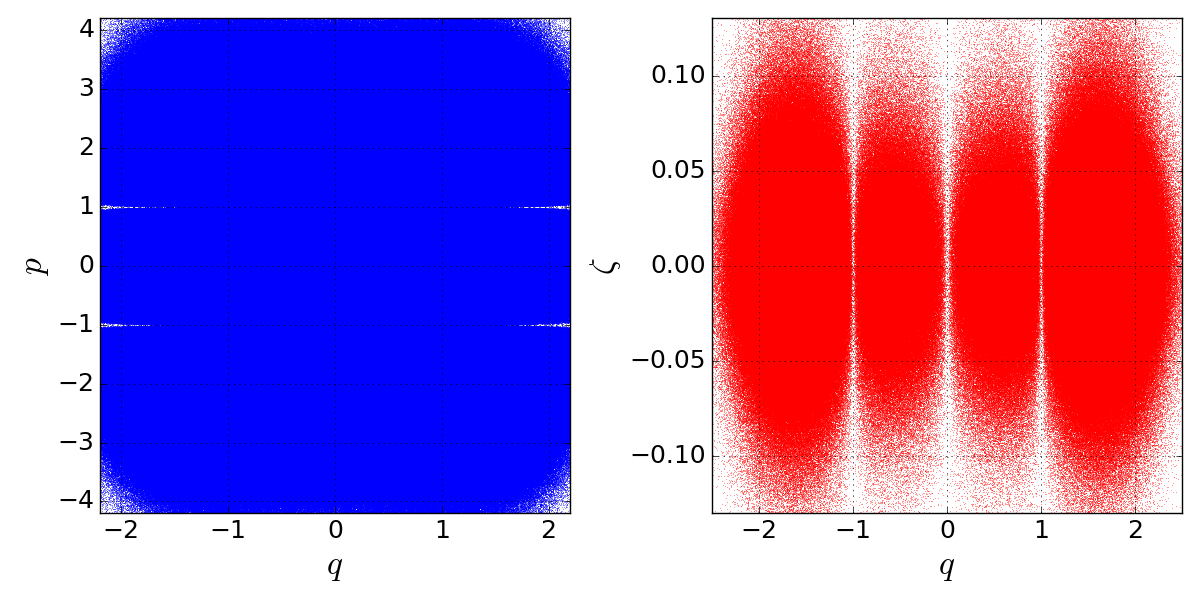}
  \caption{Poincar\'e sections for the Mexican hat potential. Around $1 \times 10^8$ crossings are shown for the section $\zeta = 0$ (left)
  {and} $4 \times 10^6$ crossings for the section $p = 0$ (right). Additionally, the nullcline lines $p = \pm 1$ (left) and $q = 0, \pm 1$ (right) are observed.}
  \label{poincaremexican}
\end{figure}

We pick a random initial condition (weighted as in the previous subsection) and integrate 
numerically the equations \eqref{logistic1}--\eqref{logistic3} using the adaptive Dormand--Prince Runge--Kutta ($4$--$5$) integrator up to a total time of $1.25 \times 10^7$. 
Then we choose two cross sections, given by $\zeta = 0$ and $p=0$ respectively, and record a point each time the section is crossed.
{In this way} we construct {the} figures \ref{poincareharmonic}, \ref{poincarequartic} and \ref{poincaremexican}. 
We visually observe the absence of ``holes'' in the cross sections, which constitutes an additional indication of  ergodicity.

\subsection{Marginal distributions}
\begin{figure}[h!]
  \centering
  \includegraphics[width=0.8\columnwidth]{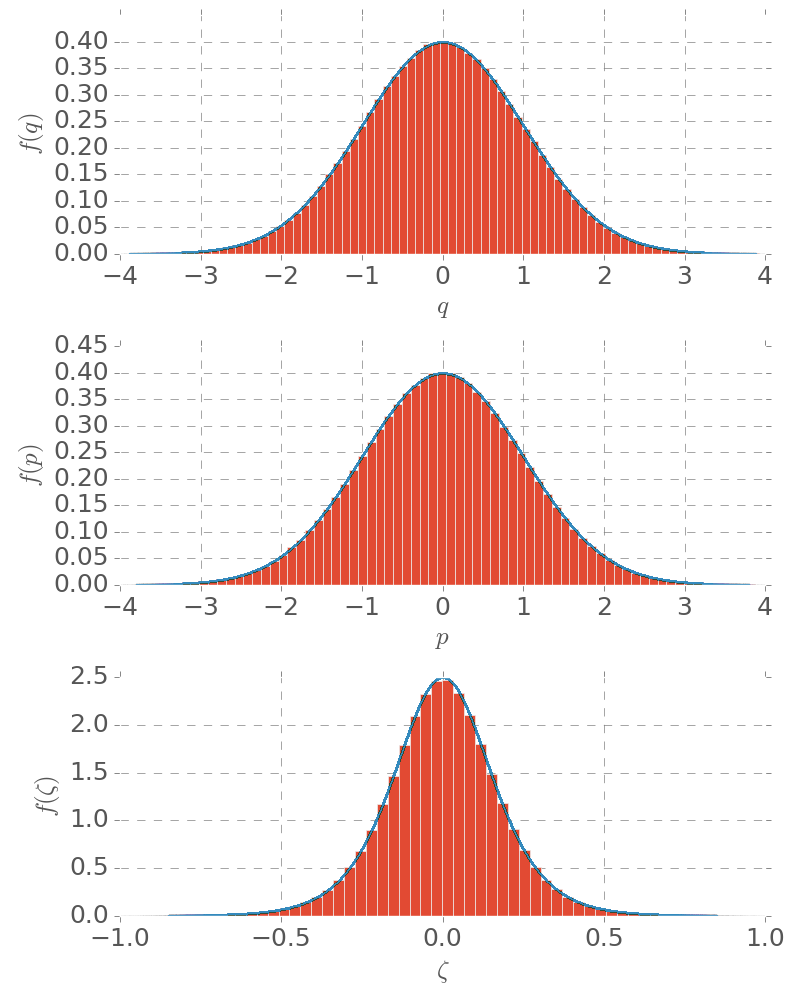}
  \caption{Histograms compared with exact marginal distributions (solid line) for the harmonic potential.}
  \label{histharmonic}
\end{figure}

\begin{figure}[h!]
  \centering
    \includegraphics[width=0.8\columnwidth]{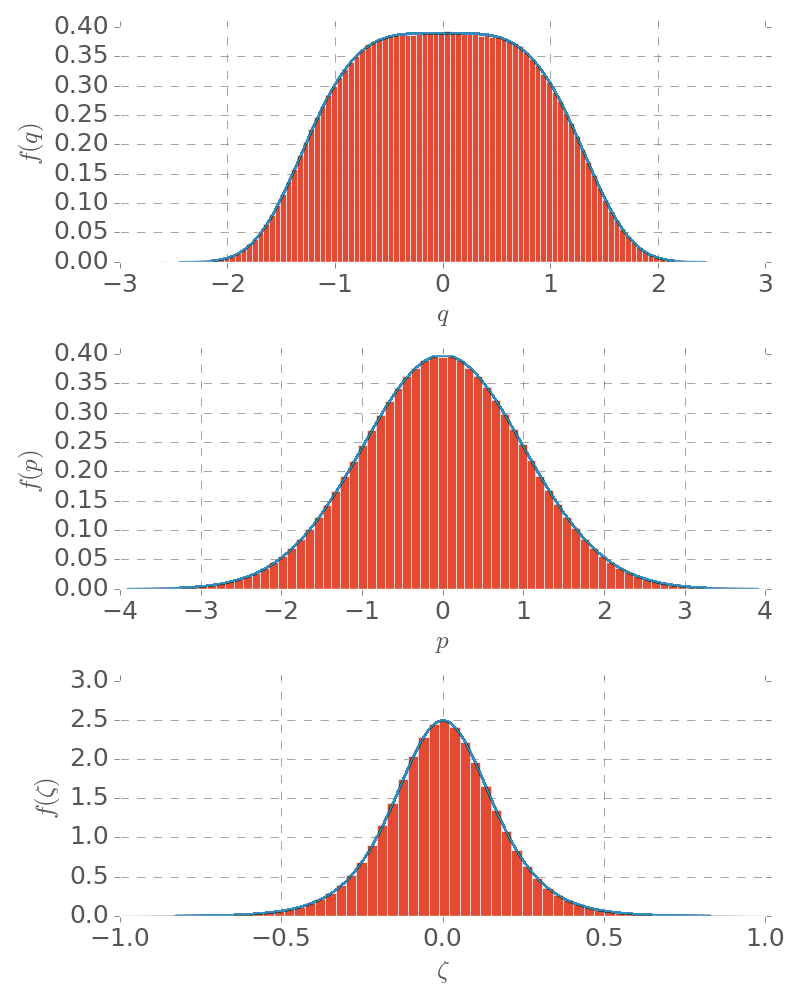}
  \caption{Histograms compared with exact marginal distributions (solid line) for the quartic potential.}
  \label{histquartic}
\end{figure}

\begin{figure}[h!]
  \centering
  \includegraphics[width=0.8\columnwidth]{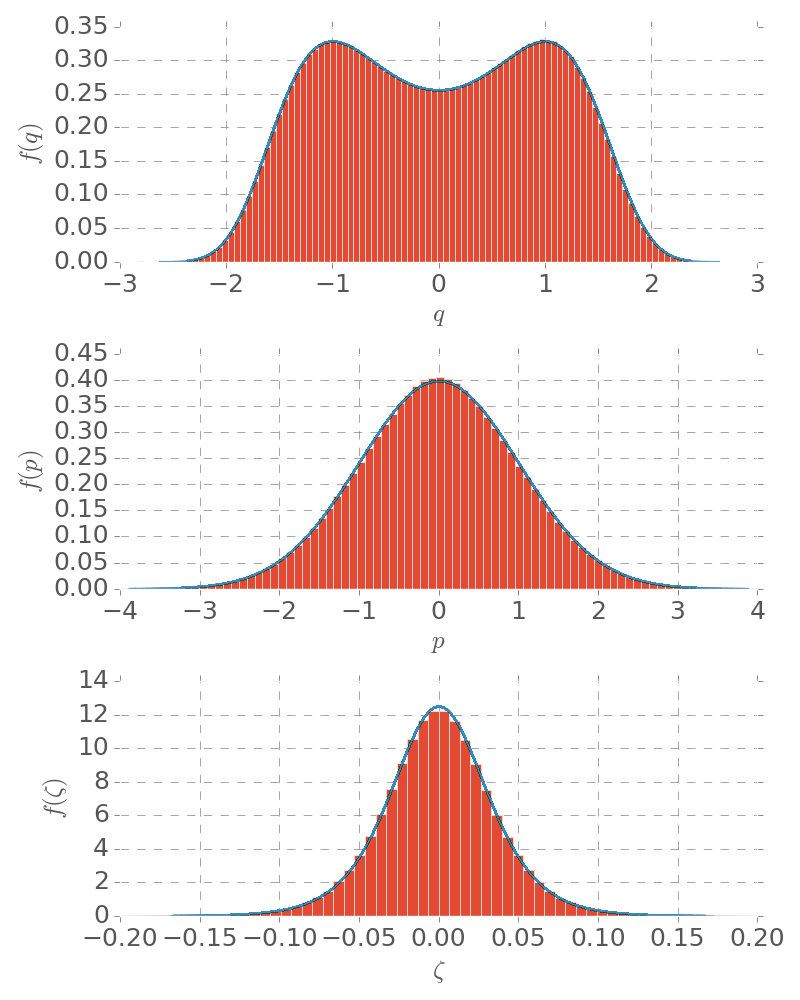}
  \caption{Histograms compared with exact marginal distributions (solid line) for the Mexican hat potential.}
  \label{histmexican}
\end{figure}

{Having determined the existence of the chaotic sea, we proceed to analyze the relation between the distributions.}
In figures \ref{histharmonic}, \ref{histquartic} and \ref{histmexican} 
we check that the {numerical} marginal distributions correspond to the theoretical ones. 
{In the next section we provide a stronger test, which confirms the convergence of the joint distribution.}

\subsection{Hellinger distance}

The DD formalism, by construction, {predicts that the joint invariant probability density is \eqref{probdensity}, 
where in our case $f(\zeta)$ is given by \eqref{logisticdistribution} and $\mu =0$. Explicitly, we have}
	\beq
	\rho(p, q, \zeta; Q) = \frac{{\rm e}^{-\beta H(q,p)}}{\cal Z}  \frac{ {\rm e}^ {\frac{\zeta}{Q}} }{ Q(1 + {\rm e}^{\frac{\zeta}{Q}})^2 } \, .
	\label{invariantdistribution}
	\eeq
{In this section} we analyze the convergence of the numerical joint distribution associated {with} a very long trajectory 
{to the theoretical} invariant distribution \eqref{invariantdistribution}. 
{For the comparison we {use a} measure of distance between distributions, 
the Hellinger distance, which
{in the extended phase space is defined as~\cite{basu2011statistical}
	\beq
	\displaystyle D_H(g || f ) = 2 \int \int \int \left(\sqrt{g} - \sqrt{f} \right)^2 \, dq \, dp \, d \zeta \, ,
	\label{hellingerdistance}
	\eeq
where $f$ and $g$ are two three-variate distributions.}
{{To calculate this distance,} we again integrate a random initial condition with the Dormand--Prince Runge--Kutta ($4$--$5$) integrator {for} 
a total time $t = 1.25 \times 10^6$ and sample $q,p,\zeta$ at a uniform time $dt_{\rm {sampling}} = 0.125$. 
{For each time interval we determine the experimental joint density by using the Kernel Density Estimation 
method \cite{basu2011statistical} and then we integrate numerically the equation~\eqref{hellingerdistance} by 
considering $g$ as the experimental density and $f$ the theoretical one~\eqref{invariantdistribution}. 
The domain of integration corresponds to the smallest rectangular domain in the extended phase space that contains the whole region explored by the trajectory.}}
The results of the {evolution of the} Hellinger distance 
with time are displayed in figure~\ref{hellingerplot}. 
\begin{figure}[t!]
  \centering
  \includegraphics[width=\columnwidth]{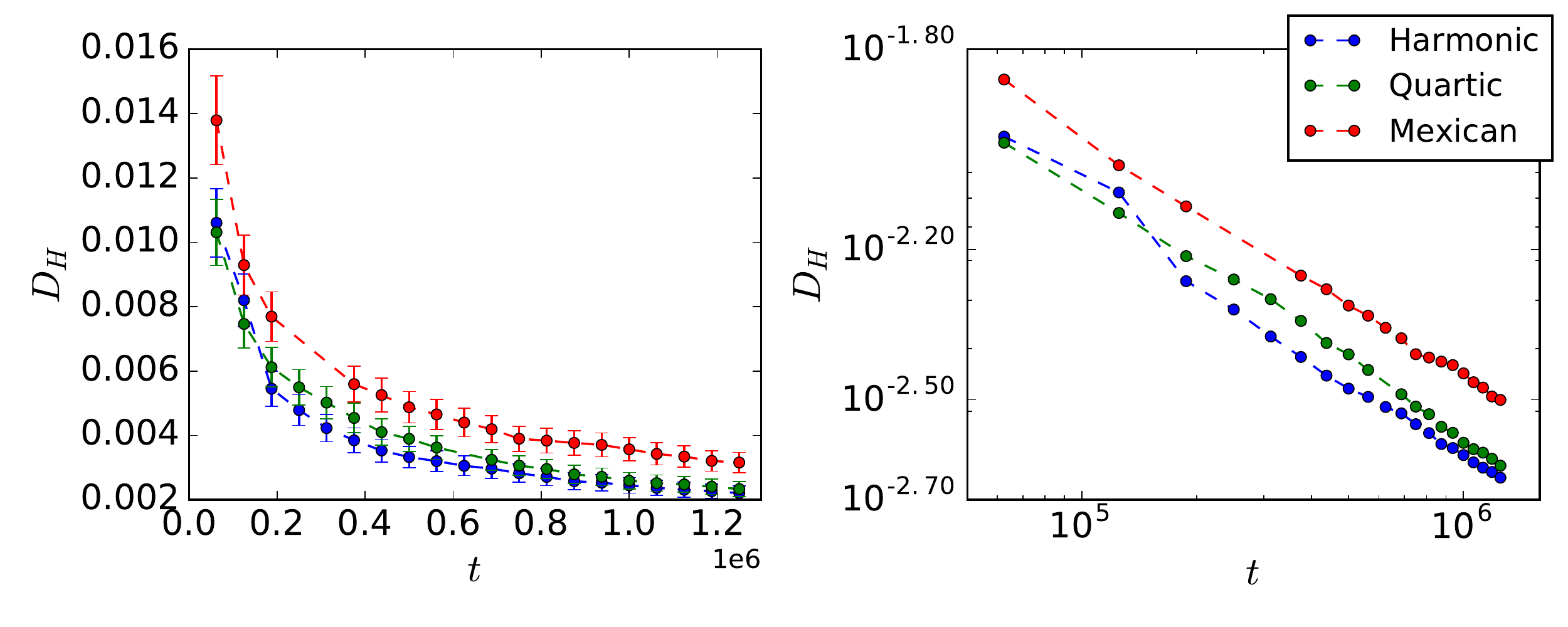}
  \caption{Hellinger distance for the three potentials as a function of integration time. The right panel shows the log-log plot.}
  \label{hellingerplot}
\end{figure}
As the figure reveals, there is a convergence to
 the expected distribution with time in {all} three cases. 
{This completes our study of ergodicity for the potentials considered.}

\section{Conclusions}
\label{conclusions}

{In this work we have performed a thorough numerical investigation on the ergodicity  
of three important singly-thermostatted one-dimensional systems.
{We employed} a logistic thermostat within the context of the Density Dynamics formalism, with the {corresponding} 
equations of motion being a set of coupled time-reversible differential equations, see~\eqref{logistic1}--\eqref{logistic3}. These equations have the same structure as those of Nos\'e--Hoover, 
but they differ in the friction term, being linear in the Nos\'e--Hoover case and  highly non-linear in our (logistic) case.}

For the one-dimensional Hamiltonian systems studied, with a quadratic, 
quartic and Mexican hat potentials, we numerically studied their ergodicity using four tests:
\begin{itemize}
\item Independence of {the} Lyapunov spectrum from the initial condition.
\item No visual holes in {the} Poincar\'e sections.
\item Agreement between marginal distributions and numerical frequencies.
\item Convergence of the joint numerical distribution to the theoretical one, quantified by the Hellinger distance.
\end{itemize}

{All the systems considered passed these numerical tests for ergodicity, thus providing strong numerical evidence that 
the dynamics of the logistic thermostat with suitable parameter values is ergodic for such systems.
The programs used for the simulations, written in the Julia language, are available at \cite{julialyapunov}.}
Our results {show the relevance of the Density Dynamics formalism as a method} to 
generate dynamics compatible with an arbitrary probability distribution. Additionally, we remark the superiority of the 
logistic thermostat {to enhance ergodicity} with respect to {other thermostats} previously used in this framework {\cite{branka2003generalization}.}  

In future work, we plan to explore in depth the structure of the phase space as the parameters $Q$ and $\beta$ are varied. 
As the ST1DS are time-reversible dynamical systems, they present characteristics which are very similar to those of Hamiltonian systems 
(e.g.~periodic orbits, tori, stochastic regions, etc.) \cite{roberts1992chaos, lamb1998time}. This structure has been analyzed, for instance, 
for the harmonic oscillator coupled to the Nos\'e--Hoover thermostat, showing very interesting properties \cite{posch1986canonical, wang2015invariant, wang2015vast}. 
An analysis of this kind may help to understand the nature of the ergodic behaviour displayed for the parameters chosen in this work.

{Additionally, it would be a challenging task to consider a theoretical approach to ergodicity of thermostatted systems by exploiting its geometric structure, as has been done for hamiltonian systems \cite{liverani1995ergodicity}.}}

\section*{Acknowledgements}
The authors thank Edison Montoya and Uriel Aceves for technical and computational support. 
AB is supported by a DGAPA-UNAM postdoctoral fellowship. DT acknowledges financial support from CONACYT, CVU No. 442828. 
DPS acknowledges financial support from DGAPA-UNAM grant PAPIIT-IN117214, and from a CONACYT sabbatical fellowship, and thanks Alan Edelman and the Julia group at MIT for hospitality while this work was carried out.

\bibliographystyle{ieeetr}
\bibliography{lyapbiblio}

\end{document}